# Subdomains of Post-COVID-Syndrome (PCS) – A Population-Based Study

Sabrina Ballhausen[1,a,*], Anne-Kathrin Ruß[2,3,a], Wolfgang Lieb[3], Anna Horn[4,5], Lilian Krist[6], Julia Fricke[6], Carmen Scheibenbogen[7], Klaus F. Rabe[8,9], Walter Maetzler[10], Corina Maetzler[10], Martin Laudien[11], Derk Frank[12], Jan Heyckendorf[1], Olga Miljukov[4,5], Karl Georg Haeusler[13], Nour Eddine El Mokthari[14], Martin Witzenrath[15], Jörg Janne Vehreschild[16,17,18], Katharina S. Appel[16,17], Irina Chaplinskaya-Sobol[19], Thalea Tamminga[1], Carolin Nürnberger[4,5], Lena Schmidbauer[4,5], Caroline Morbach[20,21], Stefan Störk[20,21], Jens-Peter Reese[4,5], Thomas Keil[4,6,22], Peter Heuschmann[4,5,23], Stefan Schreiber[1], Michael Krawczak[3], Thomas Bahmer[1,9]

[a] These authors contributed equally

1. Internal Medicine Department I, University Hospital Schleswig Holstein Campus Kiel, Kiel, Germany
2. Institute of Medical Informatics and Statistics, Kiel University, University Hospital Schleswig-Holstein Campus Kiel, Kiel, Germany
3. Institute of Epidemiology, Kiel University, University Hospital Schleswig-Holstein, Kiel, Germany
4. Institute of Clinical Epidemiology and Biometry, University of Würzburg, Würzburg, Germany
5. Institute for Medical Data Sciences, University Hospital Würzburg, Würzburg, Germany
6. Institute of Social Medicine, Epidemiology and Health Economics, Charité – Universitymedicine Berlin, Berlin, Germany

7. Institute of Medical Immunology, Charité – University medicine Berlin, Berlin, Germany

8. LungenClinic Großhansdorf, Pneumology, Großhansdorf, Germany

9. Airway Research Center North (ARCN), German Center for Lung Research (DZL), Großhansdorf, Germany

10. Neurology Department, University Hospital Schleswig Holstein Campus Kiel, Kiel, Germany

11. ENT Department, University Hospital Schleswig Holstein Campus Kiel, Kiel, Germany

12. Internal Medicine Department III, University Hospital Schleswig Holstein Campus Kiel, Kiel, Germany

13. Clinic for Neurology, University Hospital Ulm, Ulm, Germany

14. Schön Klinik Rendsburg, Cardiology, Rendsburg, Germany

15. German Center for Lung Research (DZL) and Department of Infectious Diseases and Respiratory Medicine and Critical Care, Charité – University medicine Berlin, Berlin, Germany

16. Goethe University Frankfurt, University Hospital, Center for Internal Medicine, Medical Department 2 (Hematology/Oncology and Infectious Diseases), Frankfurt on the Main, Germany

17. University of Cologne, Faculty of Medicine and University Hospital Cologne, Department I for Internal Medicine, Cologne, Germany

18. German Center for Infection Research (DZIF), partner site Bonn-Cologne, Cologne, Germany

19. Institute for Medical Informatics, University Medical Center Göttingen, Göttingen, Germany

20. Department of Clinical Research and Epidemiology, Comprehensive Heart Failure Center, University Hospital Würzburg, Würzburg, Germany

21. Department of Internal Medicine I, University Hospital Würzburg, Würzburg, Germany

22. State Institute of Health, Bavarian Health and Food Safety Authority, Erlangen, Germany

23. Clinical Trial Center Würzburg (CTC/ZKS), University Hospital Würzburg, Würzburg, Germany

*Corresponding author

Sabrina Ballhausen

Internal Medicine Department I

University Hospital Schleswig-Holstein (UKSH) Campus Kiel,

Arnold-Heller-Straße 3

24105 Kiel,

Germany

Phone: +49 (0)431 500-69471

Mail: Sabrina.ballhausen@web.de

Sabrina.Luebcker-Ballhausen@uksh.de

**ABSTRACT**

'Post-COVID Syndrome' (PCS), encompassing the multifaceted sequelae of COVID-19, can be severity-graded using a score comprising 12 different long-term symptom complexes. Acute COVID-19 severity and individual resilience were previously identified as key predictors of this score. This study validated these predictors and examined their relationship to PCS symptom complexes, using an expanded dataset (n=3,372) from the COVIDOM cohort study. Classification and Regression Tree (CART) analysis resolved the detailed relationship between the predictors and the constituting symptom complexes of the PCS score. Among newly recruited COVIDOM participants (n=1,930), the PCS score was again found to be associated with both its putative predictors. Of the score-constituting symptom complexes, neurological symptoms, sleep disturbance, and fatigue were predicted by individual resilience, whereas acute disease severity predicted exercise intolerance, chemosensory deficits, joint or muscle pain, signs of infection, and fatigue. These associations inspired the definition of two novel PCS scores that included the above-mentioned subsets of symptom complexes only. Both novel scores were inversely correlated with quality of life, measured by the EQ-5D-5L index. The newly defined scores may enhance the assessment of PCS severity, both in a research context and to delineate distinct PCS subdomains with different therapeutic and interventional needs in clinical practise.

**INTRODUCTION**

The rapid spread of SARS-CoV-2, a coronavirus strain first identified in December 2019 in Wuhan, China, resulted in a global pandemic of the associated coronavirus disease 2019 (COVID-19). COVID-19 manifests itself through a variety of symptoms, most notably respiratory complications (i.e., cough, dyspnoea, and shortness of breath) and fever [1]. If COVID-19-associated health problems persist until 12 weeks or later after the initial infection, patients are diagnosed with Post-COVID Syndrome (PCS) [2].

PCS is a complex and heterogeneous condition predominantly characterized by fatigue, exertional intolerance, and memory or concentration deficits [3,4]. Worldwide, estimates of the prevalence of PCS vary greatly depending upon the disease definition and diagnostic criteria used. In patients with mostly mild COVID-19, as are usually included in population-based follow-up studies, the prevalence of self-reported PCS was found to be approximately 15%. However, although it is known that former COVID-19 patients consult general practitioners 1.2 times more often than non-infected controls, the true public health impact of PCS is still unclear [5–7]. This uncertainty is partly due to limited treatment options and varying levels of healthcare utilization for PCS, which so far may have obscured the true rate of convalescence among COVID-19 patients.

The PCS score developed by Bahmer et al. [4] in 2022 allows individuals to be categorized as either not/mildly, moderately, or severely affected by PCS, based upon 12 self-assessed symptom complexes. Higher PCS scores were consistently found to be correlated with lower quality of life (QoL), as measured by the EQ-5D-5L index [4]. Notably, the severity of acute COVID-19 and the individual resilience of patients were consistently identified as relevant predictors of the score-based PCS severity of study participants in two different cohorts [4]. One possible explanation of this finding could be that the PCS score encompasses symptom complexes that affect patients differently, depending upon their individual resilience or severity of acute illness. The original definition of the PCS score did not account for this possibility.

A extended detailed characterization of PCS, particularly the identification of possible subdomains of the PCS phenotype, would not only be of scientific interest but could also improve PCS care by enabling more targeted treatment strategies. Therefore, we analysed a large expansion of the database used for the definition of the PCS score (i) to verify its two main predictors and (ii) to assess in increased detail how the latter relate to the symptom complexes constituting the PCS score.

**MATERIALS AND METHODS**

**The COVIDOM Study**

COVIDOM is a prospective population-based cohort study that aims to investigate the long-term health effects of COVID-19 in Germany. The study was initiated in October 2020 as part of the National Pandemic Cohort Network (NAPKON), funded by the Federal Ministry of Education and Research. In collaboration with local health care authorities in the Kiel, Würzburg and Berlin regions, a total of 3,372 participants (Kiel: 2,413; Würzburg: 523; Berlin: 436) with a positive PCR test for SARS-CoV-2 were recruited and had their data collected between 15 November 2020 and 24 January 2023. First study site visits were scheduled 6 months or later after the initial SARS-CoV-2 infection. In order to account for the local circumstances of their recruitment, the Würzburg and Berlin samples were combined in one sub-cohort (n=959), and the Kiel samples were divided into sub-cohorts Kiel-I (n=667) and Kiel-II (n=1,746). While Kiel-I coincides with its namesake in the study by Bahmer et al. [4], Würzburg/Berlin and Kiel-II represent expansions by 643 and 1,287 participants, respectively, of the original sub-cohorts. Additional information about COVIDOM has been published elsewhere [4,8].

Ethic Committee

COVIDOM has been carried out in accordance with relevant guidelines and regulations. The study was approved by the local ethic committees of the university hospitals of Kiel (No. D 537/20) and Würzburg (No. 236/20_z). According to the professional code of the Berlin Medical Association, approval by the Kiel ethics committee was also valid for the Berlin study site. All participants provided written informed consent prior to their inclusion.

Study protocol

The COVIDOM study protocol included a detailed patient history and a structured interview, followed by in-depth clinical examinations and biomaterial collection [4]. Further information about the data acquisition process can be found in the Supplementary Material of this article. In brief, before and during their study site visit, participants were asked to answer questionnaires comprising validated diagnostic tools, namely PHQ-8 for depression, GAD-7 for anxiety, FACIT-F for fatigue, screening questions according to the Canadian Consensus Criteria for 'myalgic encephalomyelitis/chronic fatigue syndrome' (ME/CFS), BRS for resilience, MoCA for cognitive function, and EQ-5D-5L for QoL, amongst others. In addition, detailed assessments were made of different organ functions (neurologic, pneumologic, cardiologic, and chemosensoric). Finally, comprehensive laboratory analyses were performed, and biomaterial was collected and stored for future analyses.

Long-term COVID-19 symptom assessment

From the collected data, the presence or absence of 35 common long-term symptoms of COVID-19 was determined for each participant. The results were summarized as binary indicators of the 12 symptom complexes originally used to define the PCS score [4], with further details provided in Table 1.

PCS score predictors

Two significant predictors of the PCS score have been identified before among COVIDOM participants, namely (i) their individual level of resilience, as measured by the Brief Resilience Scale (BRS) index, and (ii) the severity of their acute illness, as measured by the number of acute phase symptoms of COVID-19 (out of a possible 24) that they individually rated as severe or life-threatening [4]. To revisit these predictors in the present study, resilience was classified as either low (BRS<3.0), medium (3.0≤BRS<4.3), or strong (4.3≤BRS) [9], and the acute severity of COVID-19 was classified as either none (no severe or life-threatening acute phase symptoms), weak (1-3 symptoms), moderate (4-6 symptoms), or severe (≥7 symptoms).

**Statistical analysis**

Statistical analysis with R 4.1.2 (https://www.R-project.org) was performed in different subgroups of COVIDOM participants with complete information on the 12 long-term symptom complexes underlying the PCS score (total n=2,889). PCS score differences between subgroups of newly recruited participants with complete data (grouping according to resilience: n=770; acute severity: n=1,268) were tested for statistical significance using a Kruskal-Wallis test or a paired Wilcoxon test, as appropriate. Pairwise associations between the 12 symptom complexes constituting the PCS score were quantified by Cohen's κ, using R packages *DescTools* and *stats*.

CART analysis

Classification and Regression Tree (CART) analysis as implemented in R package *rpart* was used to resolve the relationship between individual resilience or acute phase COVID-19 severity, respectively, and the 12 PCS score-constituting symptom complexes. The analyses included all COVIDOM participants with complete data on both the score and the respective predictor (resilience: n=2,250; acute severity: n=2,889). While either individual resilience or acute COVID-19 severity served as the respective target variable in one of two separate CART analyses, the binary symptom complex indicators were used as potential classifiers in both instances. Node purity was measured by the Gini index, and the number of cross-validations was set to 1,000. To avoid overfitting, the resulting trees were pruned until each terminal node comprised at least 5% of the samples. Candidate symptom complexes for the subsequent construction of two novel, predictor-specific PCS scores were selected by the elbow method, drawing upon the importance values of individual symptom complexes as provided by *rpart*. In the process, importance value thresholds of 14.3 and 22.5 were employed for resilience and acute severity, respectively.

Construction of predictor-specific PCS scores

Based upon the results of the CART analysis, two novel PCS scores were constructed in sub-cohort Kiel-I (n=667) following the same procedure as in the original PCS score definition, but in predictor-specific fashion [4]. In brief, each CART-selected subset of symptom complexes was subjected to an iterative combination of k-means clustering and ordinal logistic regression analysis, treating the cluster affiliation of a participant as the respective outcome variable. Cluster number k was increased until the logistic regression models became sufficiently stable according to the Pearson correlation coefficient between the scores resulting from subsequent regression models. The regression coefficients of the final models served as weights in the subsequent PCS score definitions [4]. Thresholds for the assignment of participants to clinically meaningful subgroups were derived for both scores by receiver operator curve (ROC) analysis, as described before [4].

*Post-hoc* identification of predictors of novel PCS scores

Potential predictors of the two novel PCS scores were evaluated *post-hoc* by way of multiple ordinal logistic regression analysis with backward selection (threshold $p<0.05$) as described [4], treating the respective PCS score class (for definition, see Results) as the outcome variable. Potential predictors were chosen from the 10 acute phase and general characteristics of COVIDOM participants that had been identified before as being significantly associated with the original PCS score [4]. Missing values of predictor variables were imputed by multiple imputation assuming that missingness was completely at random. Two independent analyses were carried out in substantially expanded COVIDOM sub-cohorts Würzburg/Berlin (n=959) and Kiel-II (n=1,746), neither of which overlapped with sub-cohort Kiel-I in which the scores were constructed.

## RESULTS

**Resilience and acute COVID-19 severity as predictors of the PCS score**

Of the potential predictors tested, only individual resilience and the acute phase severity of COVID-19 were originally found by Bahmer et al. [4] to be consistently associated with the PCS score. This result was confirmed in the present study for COVIDOM participants who have since been recruited. Among new participants with complete data on both the PCS score and the BRS index (n=770), the former was found to decrease with increasing individual resilience (Figure 1a) and to differ significantly between all three BRS-defined resilience classes (Kruskal-Wallis test or paired Wilcoxon test, as appropriate: all Benjamini-Hochberg-adjusted p<0.001). Similarly, among the 1,268 newly recruited participants with complete data on the PCS score and acute COVID-19 severity, more serious illness was associated with a higher PCS score (Figure 1b) and the PCS score differed significantly between all acute severity-defined groups (Kruskal-Wallis test or paired Wilcoxon test, as appropriate: all Benjamini-Hochberg-adjusted p<0.001).

**Association between long-term symptom complexes**

Pairwise associations between the 12 long-term symptom complexes included in the PCS score were quantified by Cohen's κ in all COVIDOM participants with complete data (n=2,889). This way, a clustering became apparent of four of the symptom complexes, namely exercise intolerance, sleep disturbance, fatigue, and neurological ailments, as shown in Figure 2. The strongest associations of all were observed between fatigue and neurological ailments (κ=0.59), and between fatigue and sleep disturbance (κ=0.50).

**Classification and Regression Tree (CART) analysis**

Individual resilience

The majority (63.5%) of COVIDOM participants with complete data on both, the PCS score and the BRS index (n=2,250) had medium resilience, 15.9% described their resilience as strong, and 20.6% as low (Figure 3a). CART analysis resulted in relevant splits in the final (pruned) tree by the presence or absence of neurological ailments, sleep disturbance, and fatigue, respectively. Applying the elbow rule to the corresponding importance values, the same three symptom complexes were also selected for the subsequent construction of a novel, resilience-specific PCS score (Figure 3b).

Acute COVID-19 severity

A second CART analysis was performed treating the acute COVID-19 severity of COVIDOM participants (n=2,889) as the target variable. Overall, 22.9% of these individuals reported no severe or life-threatening symptoms, 42.7% were mildly affected, 20.5% reported moderate severity, and 13.8% suffered severely from the disease. Tree building and pruning led to four splits by the presence or absence of exercise intolerance, infection signs, chemosensory deficits, and fatigue (Figure 4a). Based upon their importance values, however, the following five PCS symptom complexes were selected for acute severity-specific score construction: exercise intolerance, fatigue, chemosensory deficits, joint or muscle pain, and infection signs (Figure 4b).

**Construction of novel predictor-specific PCS scores**

Following Bahmer et al. [4], participants in sub-cohort Kiel-I with complete data on the CART-selected symptom complexes (n=605) were included in the construction of two novel, predictor-specific PCS scores. In brief, the respective symptom complexes were subjected to an iterative combination of k-means clustering and ordinal logistic regression, each time treating the cluster affiliation of a sample as the outcome variable in the regression analysis. Parameter k was increased until the logistic regression models became sufficiently stable [4].

For individual resilience, the construction involved three long-term symptom complexes, namely neurological ailments, fatigue, and sleep disturbance. It stopped at k=2, and the resulting clusters clearly differed in terms of the proportions of participants presenting individual symptom complexes, with further details provided in Table 2. The mean cluster centre equalled 0.137 for the first cluster (n=224), and 0.893 for the second cluster (n=381). All three symptom complexes were assigned roughly the same weight in the resulting resilience-specific PCS-R score (fatigue: 46.0; neurological ailments: 46.0; sleeping disturbance: 45.0). ROC analysis led to a threshold of 46.0 (AUC=0,971) for the PCS-R score to distinguish between a severe (cluster II) and a less severe or absent PCS phenotype (cluster I). Classification according to this threshold reproduced the cluster affiliation for 592 of the 605 participants (97,9%).

For the acute disease severity-specific PCS-S score, the construction process stopped at k=6, with mean cluster centres of 0 0.007, 0.240, 0.382, 0.408, 0.635, and 0.767, respectively, with further details provided in Table 3. Joint or muscle pain and chemosensory deficits were assigned the highest weights (7.0 and 6.5, respectively), followed by exercise intolerance (4.5), infection signs (3.9) and fatigue (2.5). ROC analysis yielded thresholds for the PCS-S score of 0.0 (AUC=0.981), 6.0 (AUC=0.970), 7.0 (AUC=0.901), 9.5 (AUC=0.951) and 13.5 (AUC=0.973), application of which reproduced the cluster affiliation of 493 of the 605 participants (81.5%).

Similar to Bahmer et al. [4], we related the two novel PCS scores to the self-reported QoL of participants, measured by EQ-5D-5L [4,10]. Both scores exhibited a significant inverse correlation with the EQ-5D-5L index (PCS-R: Spearman correlation coefficient $\rho$=-0.63; PCS-S: $\rho$=-0.64; both p<0.001), indicating that higher PCS-R or PCS-S scores are associated with significantly lower QoL. Notably, however, these correlations were found to be slightly weaker than for the original PCS score ($\rho$=-0.70; p<0.001).

The PCS-R and PCS-S scores were also moderately correlated with one another ($\rho$=0.71). This notwithstanding, linear regression analysis revealed that the two novel scores, which comprise only seven of the 12 symptom complexes underlying the original PCS score, predict the latter exceptionally well. The resulting model equalled

(1)     PCS = -0.219 + 0.124×PCS-R + 1.048×PCS-S

and achieved a coefficient of determination of $R^2$=0.93, i.e., 93% of the variation of the original PCS score was explicable by the variation of the two novel scores. For practical purposes, it may be useful to rescale the two novel PCS scores so that their sum roughly equals the original score. To this end, each logistic regression coefficient would have to be multiplied by the corresponding linear regression coefficient and the result rounded again to the nearest half-integer (see Tables 2 and 3).

### *Post-hoc* identification of predictors of novel PCS scores

To assess the robustness of their intended predictor specificity, the two novel PCS scores were subjected to multiple ordinal logistic regression analyses with backward selection (threshold: p<0.05) in COVIDOM sub-cohorts Würzburg/Berlin (n=884) and Kiel-II (n=1,613), treating the corresponding PCS score classes as the respective outcome variables.

In addition to individual resilience itself, only the most severe form of acute COVID-19 was found to be consistently associated with a higher PCS-R score (Supplementary Table 1). While a one-unit increase in BRS index reduces the odds for a high PCS-R score by 37% to 57%, seven or more serious or life-threatening symptoms during the acute phase of COVID-19 increased these odds by a factor of 2.7 to 4.5, compared to a lack of symptoms.

The PCS-S score was also associated with acute disease severity (Supplementary Table 2), and the estimated odds ratios of the three higher severity categories (i.e. 1-3, 4-6, ≥7 symptoms), relative to the lack of symptoms, ranged from 1.9 to 3.2 in the Würzburg/Berlin sub-cohort, and from 2.9 to 10.5 in the Kiel-II sub-cohort. As the only other consistent predictor in Kiel-II and Würzburg/Berlin, a one-unit increase in body mass index increased the odds for a high PCS-S score by 4% to 7%.

**DISCUSSION**

Although the global COVID-19 pandemic apparently has lost much of its threat, there is still a continuing need to study the long-term consequences of SARS-CoV-2 infection. Such research is important not only because of the societal and economic burden of PCS, but also to enhance preparedness for similar public health challenges in the future. However, the clinical definition of PCS remains imprecise, and symptom-based criteria that allow a classification of patients according to their medical needs are either vague or lacking altogether. In this study, we reported that the previously developed PCS score, which uses 12 different long-term symptom complexes to measure individual PCS severity [4], in fact encompasses two subdomains of the syndrome that can be captured by two novel, more specific PCS scores. Instead of covering the whole spectrum of possible sequelae of COVID-19, these scores comprise subsets of symptom complexes that were selected on the basis of their association with the two main predictors of the original PCS score, namely individual resilience and acute phase severity of COVID-19. The novel scores, labelled PCS-R and PCS-S, may not only complement the original score when tailoring interventional strategies for PCS, but potentially also improve our understanding of the pathogenesis of this highly impairing condition.

The present study confirmed the capability of individual resilience and acute COVID-19 severity to predict PCS severity and suggests that the two factors act upon the PCS phenotype through different etiological pathways. Resilience, a general trait of individuals, was found to be negatively associated with post-acute fatigue, neurological ailments, and sleep disturbances, all of which were also strongly associated with one another. The PCS-R score derived from these symptom complexes classifies patients into two severity groups, roughly speaking those with or without a condition that is independent of the acute phase severity of COVID-19 and which may be termed 'neuro-psychological' PCS. In contrast, the PCS-S score, which comprises exercise intolerance, fatigue, joint or muscle pain, chemosensory deficits, and signs of infection, defines six classes of an apparent 'prolonged recovery' subdomain of PCS.

The original and the novel PCS scores were developed as tools to classify disease severity based upon a broad spectrum of disease sequelae, not least to provide an appropriate outcome measure for clinical trials. Other studies pursued different strategies when devising PCS scoring systems or algorithms for PCS sub-phenotype identification. For instance, the PASC score developed in the RECOVER study was intended for the differential diagnosis of PCS [11]. Symptoms more specific for PCS, such as the impairment of smell and taste, consequently have greater weight in the PASC score than non-specific symptoms like fatigue. Although the PASC score uses binary indicators of self-reported symptoms similar to our scores, and although higher PASC scores are associated with lower health-related QoL as well, the main purpose of the PASC score was to determine whether or not a patient has PCS at all, and not to measure PCS intensity. Nevertheless, some of the score-defined groupings of patients are similar in both systems. For example, cluster IV of our PCS-S score and cluster I of the PASC score mainly comprise patients with chemosensory deficits, corroborating that these clusters may highlight a distinct phenotype [11]. Joint or muscle pain, on the other hand, is a rather frequent symptom in all PASC clusters but prevails only in cluster VI of our PCS-S score [11]. Notably, however, the PASC score also highlights the limitations of defining PCS by clinical symptoms alone because as many as 3.7% of the non-infected participants of the RECOVER study were PCS-positive, according to the PASC score [11].

In the ORCHESTRA study [12], different PCS phenotypes were also defined by way of clustering, but without prior stratification by potential PCS predictors as was done for constructing the PCS-R and PCS-R scores. The four clusters detected in ORCHESTRA were (i) chronic fatigue-like syndrome, with fatigue, headache, and memory loss; (ii) respiratory syndrome, with cough and dyspnoea; (iii) chronic pain syndrome, with arthralgia and myalgia, and (iv) neurosensory syndrome, with alterations of taste and smell. Except for respiratory syndrome, most of the characteristics of the other clusters recall the symptom complexes included in our two novel PCS scores. In line with other reports [13], female sex was also identified in ORCHESTRA as a predictor of severe PCS [12,13]. However, even although we observed a trend towards higher PCS score values in women, both for the original and the novel scores, sex was excluded as a relevant covariate from all logistic regression models during score construction when individual resilience was included.

Interestingly, a history of psychiatric diseases, especially depression, strongly predicted chronic manifestation of COVID-19 symptoms in a Swiss study [7]. In the Würzburg/Berlin sub-cohort of COVIDOM, pre-existing neurologic or psychiatric diseases were also found to be a highly significant predictor of the PCS-S score, but the same association failed to reach statistical significance in Kiel-II after adjustment for multiple testing. Together, these results suggest a possible, albeit milder effect of a neuro-psychiatric disease history in the COVIDOM data. The Swiss study also revealed how the presence and severity of PCS interferes with the ability to work and of the 1.6% of PCS patients who were unfit to work at all, the majority were women [7]. The Swiss data together with our own therefore suggest that the resilience-specific PCS subdomain identified in here shares risk factors with depression and other psychiatric diseases, both of which are also more frequent in women.

Many studies exploring risk factors and sub-phenotypes of PCS were based upon electronic health records or hospital records, and often used the International Classification of Disease (ICD) code U09.9 as a criterion for data extraction [14–16]. This has limited such studies to patients (i) who sought medical help for comparatively severe PCS and (ii) whose treating physicians actually used the ICD code in question. An example of this is provided by a US American study [15] in which an initial database of roughly 14 million patients was narrowed down to just 6,500 participants who fulfilled the inclusion criteria. Similar to our own results, the US study identified a neuropsychiatric and a pain/fatigue cluster in addition to two multi-systemic clusters [15], thereby adding further evidence for the existence of at least two distinct subdomains of PCS. However, whereas pulmonary and cardiovascular clusters were observed in the US study as well, the latter systematically excluded data on chemosensory deficits [15] because these are rarely documented in hospital records.

The strengths of the COVIDOM study include its population-based, multi-centre and prospective setting as well as its structured collection of quality-controlled data. In the future, the detailed clinical phenotyping in COVIDOM will also allow clinical classifications to be related to molecular markers as measured in the collected biomaterials. A limitation of the study is its retrospective acquisition of clinical data from the acute phase of COVID-19, potentially introducing recall bias. Furthermore, the majority of the COVIDOM data came from non-vaccinated individuals who got infected during the pre-Omicron era. Therefore, the conclusions drawn in here may not be readily applicable to other variants of SARS-CoV-2, or to vaccinated individuals. Finally, one limitation of COVIDOM may be its lack of pre-infection data. Since fatigue, sleep disturbance, and cognitive impairment are frequent symptoms of psychiatric and neurologic diseases, it cannot be excluded that the recruitment of such patients into COVIDOM may have resulted in overestimating the role of these symptoms in PCS.

In conclusion, individuals with low resilience seem to suffer differently from PCS than individuals with severe acute COVID-19. The two novel PCS scores developed above should allow clinicians and researchers to take this difference into account. Since functional limitations are scarce in PCS patients, clinical trials of PCS mostly employ patient-reported outcomes or health-related QoL to evaluate the efficacy of interventions and treatments. The proposed PCS scores can transform this information into clear-defined PCS (sub-)phenotypes, thereby enabling more efficient tailoring of interventions to different aetiologies and clinical needs.

**STATEMENTS AND DECLARATIONS**

**Acknowledgements**

We are thankful to all the staff at the study sites Kiel, Würzburg and Berlin for their excellent support. We also thank the supporting infrastructures of the NUM NUKLEUS: ECU (Epidemiology Core Unit), BCU (Biobanking Core Unit), CDM (Clinical Data Management), THS (Trusted Third Party), and ICU (Interaction Core Unit).

**Funding**

The COVIDOM study is part of the National Pandemic Cohort Network (NAPKON). NAPKON is funded by the German Federal Ministry for Education and Research (BMBF) and is administrated by the Network University Medicine (NUM). The BMBF support code of NAPKON is 01KX2021. The Kiel and Würzburg study sites are partly supported by funding from the federal states (´Länder´) of Schleswig-Holstein and Bavaria, respectively.

No funding was received to assist with the preparation of this manuscript.

**Data Availability**

The data used in this study are not freely available due to legal restrictions and limitations imposed by participant consent. The data can be obtained upon reasonable request from the NAPKON Data Use and Access Committee. For relevant data governance information and the submission of requests, visit https://proskive.napkon.de.

**LEGENDS**

**Table 1:** Long-term COVID-19 symptom complexes underlying Post-COVID Syndrome (PCS) score definition.

| Symptom complex | Self-reported sub-symptoms |
|---|---|
| Fatigue | Fatigue |
| Coughing, wheezing | Coughing, wheezing |
| Neurological ailments | Confusion, vertigo, headache, motor deficits, sensory deficits, numbness, tremor, deficits of concentration, cognition or speech |
| Joint or muscle pain | Muscle pain, joint pain |
| Ear-Nose-Throat ailments | Hoarseness, sore throat, running nose |
| Gastrointestinal ailments | Stomach pain, diarrhoea, vomiting, nausea |
| Sleep disturbance | Insomnia, unrestful sleep |
| Exercise intolerance | Shortness of breath, reduced exercise capacity |
| Infection signs | Chills, fever, general sickness/flu-like symptoms |
| Chemosensory deficits | Smelling disturbance, impaired sense of taste |
| Chest pain | Chest pain |
| Dermatological ailments | Hair loss, rash, itchiness |

Note: A symptom complex was considered present if the participant reported the presence of at least one of its sub-symptoms.

**Table 2:** Definition of resilience-specific Post-COVID Syndrome score (PCS-R score)

| No. | Symptom complex | Cluster centre I (n=224) | Cluster centre II (n=381) | Logistic regression coefficient | Score weight original | Score weight rescaled |
|---|---|---|---|---|---|---|
| 2 | Fatigue | 0.094 | 0.906 | 45.751 | 46.0 | 5.5 |
| 9 | Neurological ailments | 0.138 | 0.934 | 45.761 | 46.0 | 5.5 |
| 12 | Sleep disturbance | 0.179 | 0.840 | 45.243 | 45.0 | 5.5 |
| Mean cluster centre | | 0.137 | 0.893 | NA | NA | |

**Table 3:** Definition of acute disease severity-specific Post-COVID Syndrome score (PCS-S score)

| No. | Symptom complex | Cluster centre | | | | | | Logistic regression coefficient | Score weight | |
| --- | --- | --- | --- | --- | --- | --- | --- | --- | --- | --- |
| | | I (n=187) | II (n=119) | III (n=111) | IV (n=99) | V (n=34) | VI (n=55) | | original | rescaled |
| 2 | Fatigue | 0.000 | 1.000 | 0.910 | 0.596 | 0.971 | 0.982 | 2.421 | 2.5 | 2.5 |
| 3 | Exercise intolerance | 0.000 | 0.000 | 1.000 | 0.384 | 1.000 | 0.945 | 4.454 | 4.5 | 4.5 |
| 4 | Joint or muscle pain | 0.005 | 0.059 | 0.000 | 0.04 | 0.000 | 1.000 | 7.228 | 7.0 | 7.5 |
| 1 | Chemosensory deficits | 0.000 | 0.000 | 0.000 | 1.000 | 0.206 | 0.382 | 6.402 | 6.5 | 6.5 |
| 11 | Infection signs | 0.032 | 0.143 | 0.000 | 0.020 | 1.000 | 0.527 | 3.747 | 3.5 | 4.0 |
| | Mean cluster centre | 0.007 | 0.240 | 0.382 | 0.408 | 0.635 | 0.767 | NA | NA | |

**Figure 1:** Relationship between PCS score and individual resilience (a) and acute COVID-19 severity (b) in newly recruited COVIDOM participants with complete data (resilience: n=770; acute severity: n=1,268). Individual resilience was classified as low (BRS<3.0), medium (3.0≤BRS< 4.3), or strong (4.3≤BRS). Acute COVID-19 severity was classified as none (no severe or life-threatening symptoms), weak (1-3 symptoms), moderate (4-6 symptoms), or severe (≥7 symptoms). Both omnibus and all pairwise tests revealed significant score differences between the predictor-defined groups (Kruskal-Wallis test or paired Wilcoxon test, as appropriate; ****: Benjamini-Hochberg-adjusted p<0.001).

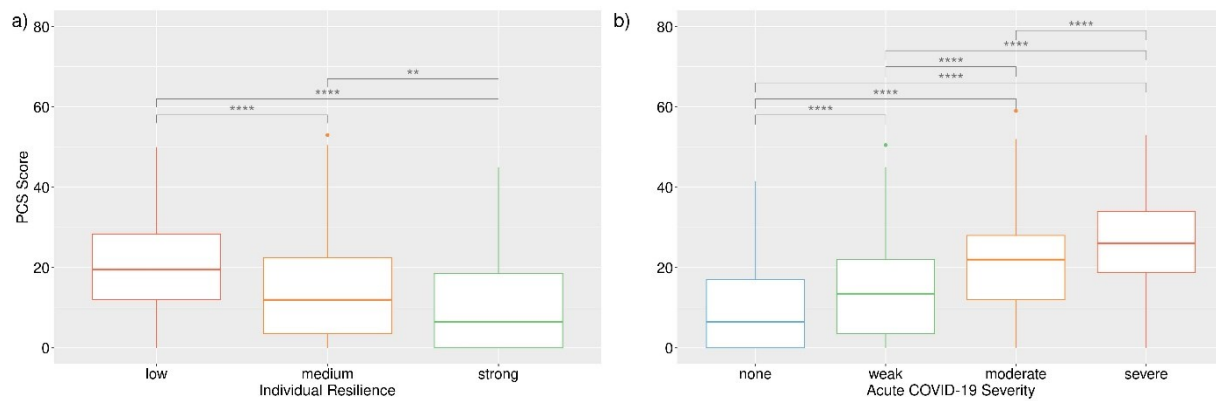

**Figure 2** Pair-wise associations between the 12 long-term symptom complexes underlying the Post-COVID Syndrome (PCS) score. Cohen's κ values were calculated in all COVIDOM participants with complete data (n=2,889) and subjected to hierarchical clustering by Euclidean distance. The strength of the association between symptom complexes was color-coded as red (0.75≤κ), yellow (0.45≤κ<0.75) or blue (κ<0.45).

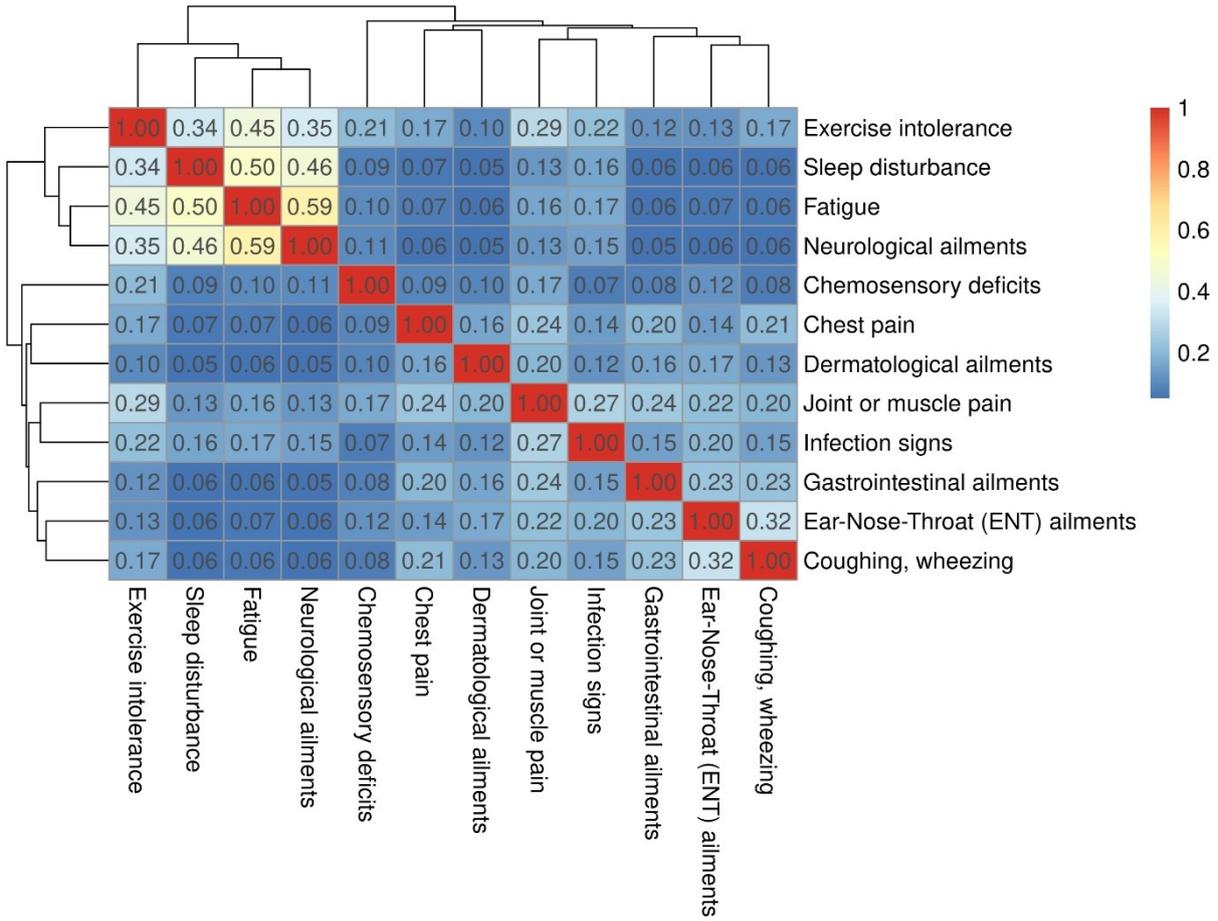

**Figure 3** Classification and Regression Tree (CART) analysis of individual resilience, as measured by the Brief Resilience Scale (BRS) index. CART analysis served to determine which PCS symptom complexes were most strongly associated with individual resilience in all COVIDOM participants with complete data (n=2,250), allowing for a possible statistical interaction between complexes. a) Nodes were successively split by the presence (1, blue) or absence (0, white) of a particular symptom complex. The percentage of samples included is given alongside each node. The resulting tree was pruned until all leaves comprised at least 5% of the samples. b) Symptom complexes were selected for novel PCS score definition by applying the elbow rule to their importance values (threshold marked by red triangle).

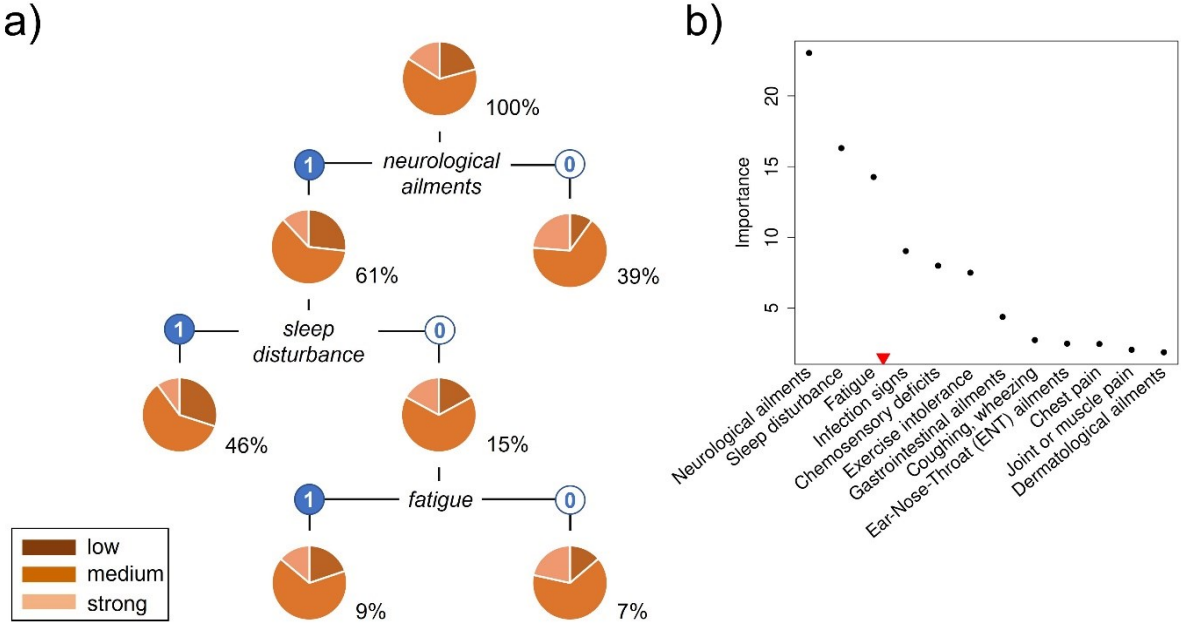

**Figure 4** Classification and Regression Tree (CART) analysis of acute COVID-19 severity (n=2,889; for details, see legend to Figure 3).

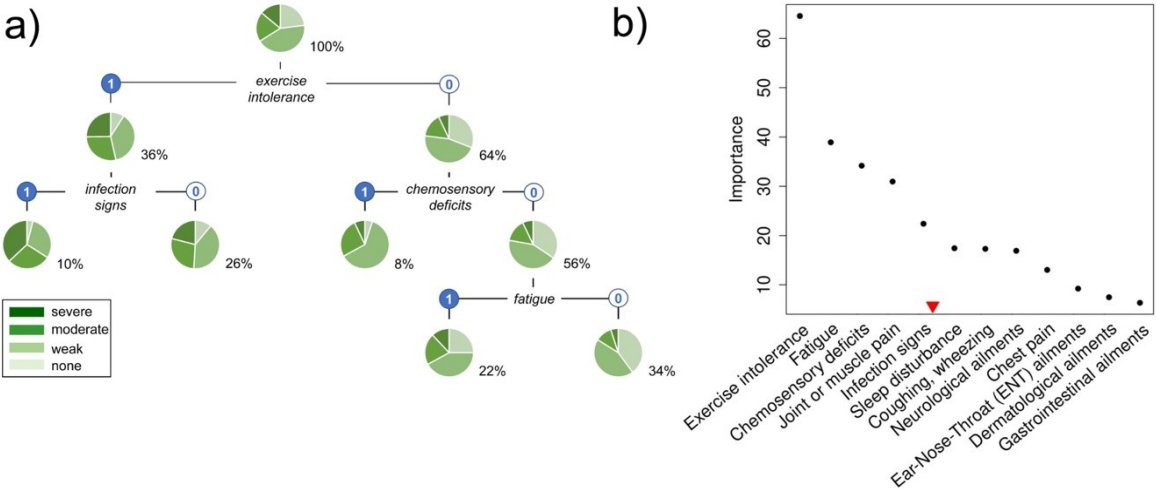

# SUPPLEMENTARY MATERIAL

**COVIDOM data acquisition**

Before visiting a study center, COVIDOM participants were asked to answer and return a questionnaire, either paper-based or online. In addition to basic personal, demographic and medical characteristics, the questionnaire included questions about the circumstances of the acute SARS-CoV-2 infection, the subsequent course of disease, and the lifestyle and current health status of the patient, particularly the persistence of COVID-19 symptoms.

During the visit to the study center, patients were examined clinically and an additional anamnesis was taken, drawing upon various validated questionnaires such as PHQ-8 (depression), GAD-7 (anxiety disorder), FACIT-F and CFS (fatigue), BRS (resilience), MoCA (cognitive function), PSS (stress), mMRC and MDP (dyspnea) as well as EQ-5D-5L (quality of life). Diagnostic measures included the assessment of body height and weight, bioelectric impedance scanning, and the control of patient vitality parameters. More specific tests were carried out from different medical disciplines, including ear-nose-throat (chemosensory and endoscopy), neurology (neurological examinations), pneumology (standardized spirometry, including body plethysmography, and diffusion capacity), cardiology (indirect blood pressure measurement, ECG, and echocardiography), hepatology (hepatic sonography and elastography), and geriatrics (6-minute walking test for probands >65 years of age).

After a final discussion with their doctor, every patient received a medical report containing their test results and, if necessary, individual health-related recommendations. All data collected at the COVIDOM study centers in Berlin, Kiel and Würzburg were deposited in a central database.

**Supplementary Table 1: Significant predictors of resilience-specific PCS-R score class in sub-cohorts Würzburg/Berlin and Kiel-II**

| Predictor variable | Level | Regression coefficient | | | Odds ratio | | P value[d] | |
|---|---|---|---|---|---|---|---|---|
| | | Estimate | Standard error | 95% confidence interval | Estimate | 95% confidence interval | Unadjusted | Adjusted |
| *Würzburg/Berlin (n=884)\** | | | | | | | | |
| Resilience (BRS) | scale | -0.460 | 0.126 | [-0.707; -0.213] | 0.631 | [0.493; 0.808] | 0.00032 | 0.0038 |
| No. serious or life-threatening symptoms[a] | 1-3 | 0.203 | 0.252 | [-0.291; 0.697] | 1.225 | [0.748; 2.007] | 0.42 | n.a. |
| | 4-6 | 0.548 | 0.318 | [-0.076; 1.171] | 1.729 | [0.927; 3.226] | 0.091 | n.a. |
| | >7 | 0.991 | 0.335 | [0.333; 1.648] | 2.693 | [1.395; 5.198] | 0.0036 | 0.043 |
| Pre-existing neurologic or psychiatric disease | yes | 1.025 | 0.234 | [0.566; 1.483] | 2.786 | [1.761; 4.407] | <0.0001 | <0.0001 |
| General anxiousness | yes | 0.517 | 0.17 | [0.184; 0.850] | 1.677 | [1.203; 2.340] | 0.0024 | 0.029 |
| Pre-existing cardiovascular disease | yes | 0.697 | 0.228 | [0.251; 1.143] | 2.008 | [1.285; 3.136] | 0.0028 | 0.034 |
| No. symptoms[b] | 3-5 | -1.034 | 0.479 | [-1.974; -0.095] | 0.355 | [0.139; 0.909] | 0.032 | 0.38 |
| | 6-8 | 0.408 | 0.371 | [-0.319; 1.135] | 1.504 | [0.727; 3.113] | 0.27 | n.a. |
| | >9 | 0.928 | 0.356 | [0.231; 1.625] | 2.531 | [1.260; 5.08] | 0.0092 | 0.11 |
| Pre-existing gastrointestinal diseases | yes | 0.809 | 0.275 | [0.269; 1.349] | 2.247 | [1.309; 3.855] | 0.011 | 0.13 |
| Body mass index | scale | 0.041 | 0.018 | [0.006; 0.075] | 1.041 | [1.006; 1.078] | 0.024 | 0.28 |

**Supplementary Table 1 (continued)**

| | | | | | | | | |
|---|---|---|---|---|---|---|---|---|
| *Kiel-II (n=1,613)\** | | | | | | | | |
| Resilience (BRS) | scale | -0.852 | 0.131 | [-1.108; -0.596] | 0.427 | [0.330; 0.551] | <0.0001 | <0.0001 |
| No. serious or life-threatening symptoms[a] | 1-3 | 0.342 | 0.187 | [-0.025; 0.708] | 1.407 | [0.975; 2.031] | 0.069 | 0.97 |
| | 4-6 | 0.803 | 0.226 | [0.359; 1.247] | 2.232 | [1.432; 3.478] | 0.00043 | 0.0060 |
| | >7 | 1.512 | 0.293 | [0.937; 2.087] | 4.535 | [2.553; 8.057] | <0.0001 | <0.0001 |
| Body weight change after infection[c] | loss | -0.463 | 0.217 | [-0.889; -0.037] | 0.629 | [0.411; 0.964] | 0.034 | 0.47 |
| | none | -0.683 | 0.185 | [-1.047; -0.320] | 0.505 | [0.351; 0.726] | 0.00049 | 0.0069 |
| Sex | male | -0.439 | 0.135 | [-0.704; -0.174] | 0.645 | [0.495; 0.840] | 0.0013 | 0.019 |
| Pre-existing neurologic or psychiatric disease | yes | 0.478 | 0.171 | [0.142; 0.814] | 1.612 | [1.152; 2.256] | 0.0055 | 0.077 |
| Gastrointestinal diseases | yes | 0.747 | 0.278 | [0.201; 1.292] | 2.110 | [1.223; 3.640] | 0.0086 | 0.12 |
| No. symptoms[b] | 3-5 | -0.487 | 0.370 | [-1.212; 0.237] | 0.614 | [0.298; 1.267] | 0.19 | n.a. |
| | 6-8 | -0.058 | 0.335 | [-0.715; 0.599] | 0.944 | [0.489; 1.821] | 0.86 | n.a. |
| | >9 | 0.808 | 0.324 | [0.173; 1.443] | 2.244 | [1.189; 4.235] | 0.013 | 0.18 |
| Body mass index | scale | 0.028 | 0.013 | [0.003; 0.053] | 1.028 | [1.003; 1.054] | 0.031 | 0.43 |
| Pre-existing cardiovascular disease | yes | 0.356 | 0.170 | [0.024; 0.689] | 1.428 | [1.024; 1.992] | 0.038 | 0.53 |

No.: number. *Includes only participants with complete data on PCSS-R. Reference levels: $^a$no symptoms, $^b$0-2 symptoms, $^c$weight gain. $^d$P values were Bonferroni-adjusted by multiplication with the total number of predictor variables present in each sub-cohort-specific regression model. Only variables for which at least one level yielded an adjusted p value <0.05 are shown. Variables with an adjusted p value <0.05 in both sub-cohorts are highlighted in green.

**Supplementary Table 2: Significant predictors of acute severity-specific PCS-S score class in sub-cohorts Würzburg/Berlin and Kiel-II**

| Predictor variable | Level | Regression coefficient | | | Odds ratio | | P[e] value | |
|---|---|---|---|---|---|---|---|---|
| | | Estimate | Standard error | 95% confidence interval | Estimate | 95% confidence interval | Unadjusted | Adjusted |
| *Würzburg/Berlin (n=884)\** | | | | | | | | |
| Body mass index | scale | 0.065 | 0.018 | [0.031; 0.1] | 1.068 | [1.031; 1.105] | 0.00023 | 0.0025 |
| No. serious or life-threatening symptoms[a] | 1-3 | 0.636 | 0.234 | [0.179; 1.094] | 1.890 | [1.196; 2.987] | 0.0081 | 0.089 |
| | 4-6 | 0.808 | 0.287 | [0.246; 1.37] | 2.243 | [1.279; 3.933] | 0.0051 | 0.056 |
| | >7 | 1.161 | 0.338 | [0.499; 1.822] | 3.192 | [1.647; 6.186] | 0.00065 | 0.0071 |
| Pre-existing neurologic or psychiatric disease | yes | 1.047 | 0.24 | [0.576; 1.518] | 2.85 | [1.779; 4.565] | <0.0001 | <0.0001 |
| No. symptoms[b] | 3-5 | -0.265 | 0.444 | [-1.135; 0.605] | 0.767 | [0.321; 1.832] | 0.55 | n.a. |
| | 6-8 | 0.391 | 0.395 | [-0.384; 1.166] | 1.478 | [0.681; 3.208] | 0.33 | n.a. |
| | >9 | 1.148 | 0.389 | [0.386; 1.911] | 3.153 | [1.471; 6.758] | 0.0045 | 0.049 |
| General anxiousness | yes | 0.418 | 0.17 | [0.085; 0.75] | 1.519 | [1.089; 2.117] | 0.014 | 0.15 |
| Age | scale | 0.013 | 0.006 | [0.002; 0.024] | 1.013 | [1.002; 1.024] | 0.018 | 0.20 |

**Supplementary Table 2 (continued)**

| | | | | | | | | |
|---|---|---|---|---|---|---|---|---|
| *Kiel-II (n=1,613)** | | | | | | | | |
| No. serious or life-threatening symptoms[a] | 1-3 | 1.046 | 0.157 | [0.738; 1.355] | 2.848 | [2.092; 3.875] | <0.0001 | <0.0001 |
| | 4-6 | 2.08 | 0.213 | [1.662; 2.497] | 8.001 | [5.27; 12.148] | <0.0001 | <0.0001 |
| | >7 | 2.352 | 0.278 | [1.807; 2.898] | 10.508 | [6.09; 18.131] | <0.0001 | <0.0001 |
| Body mass index | scale | 0.04 | 0.014 | [0.013; 0.066] | 1.040 | [1.013; 1.069] | 0.0040 | 0.044 |
| Resilience (BRS) | scale | -0.573 | 0.105 | [-0.779; -0.368] | 0.564 | [0.459; 0.692] | <0.0001 | <0.0001 |
| Sex | male | -0.472 | 0.129 | [-0.725; -0.22] | 0.624 | [0.484; 0.803] | 0.00025 | 0.0028 |
| Education[c] | university entrance | -0.391 | 0.14 | [-0.665; -0.118] | 0.676 | [0.514; 0.889] | 0.0058 | 0.063 |
| Pre-existing pulmonary disease | yes | 0.596 | 0.219 | [0.167; 1.026] | 1.815 | [1.181; 2.789] | 0.0098 | 0.11 |
| Pre-existing rheumatic disease | yes | 0.694 | 0.28 | [0.145; 1.242] | 2.002 | [1.156; 3.464] | 0.014 | 0.16 |
| Body weight change after infection[d] | loss | -0.154 | 0.225 | [-0.596; 0.287] | 0.857 | [0.551; 1.333] | 0.49 | n.a. |
| | none | -0.375 | 0.169 | [-0.706; -0.044] | 0.687 | [0.494; 0.957] | 0.027 | 0.30 |

No.: number. *Includes only participants with complete data on PCSS-S. Reference levels: [a]no symptoms, [b]0-2 symptoms. [c]other school degree, [d]weight gain. [e]P values were Bonferroni-adjusted by multiplication with the total number of predictor variables present in each sub-cohort-specific regression model. Only variables for which at least one level yielded an adjusted p value <0.05 are shown. Variables with an adjusted p value <0.05 in both sub-cohorts are highlighted in green.